# Observation of Skyrmion Bubbles in Multilayer [Pt/Co/Cu]$_n$ using spin-polarized STM

Jacob J. Repicky, Brad Goff, Shuyu Cheng, Roland K. Kawakami, Jay A. Gupta

*Department of Physics, The Ohio State University, Columbus, OH 43210, USA*

**Abstract**

Magnetic multilayers are a promising platform for storage and logic devices based on skyrmion spin textures, due to the large materials phase space for tuning properties. Epitaxial superlattice structures of [Pt/Co/Cu]$_n$ thin films were grown by molecular beam epitaxy at room temperature. Spin-polarized scanning tunneling microscopy (SP-STM) of these samples was used to probe the connection between surface structure and skyrmion morphology with nanoscale spatial resolution. Irregular-shaped skyrmion bubbles were observed, with effective diameters from 20-200 nm that are much larger than the nanoscale grain structure of the surface topography. Nucleation, annihilation, and motion of skyrmion bubbles could be driven using the stray field of the ferromagnetic tip in repeated imaging, and spin-polarized current/voltage pulses. Our detailed comparison of STM topography and differential conductance images shows that there are no surface defects or inhomogeneities at length scales that could account for the range in skyrmion bubble size or shape observed in the measurements.

Magnetic skyrmions are vortex-like configurations of magnetization with non-trivial topological character, and are stabilized by the chiral Dzyaloshinskii-Moriya interaction (DMI). These magnetic textures are of interest for use as information carriers in next-generation spin-based memory and logic devices [1–3]. Recent efforts to make such devices useful and competitive have focused on the realization of nanoscale skyrmions at room temperature and zero applied magnetic field, with desirable dynamic properties. This is a complex materials optimization problem that may be met in thin film heterostructure platforms including oxides, heavy metal/ferrimagnetic insulator bilayers, and metallic multilayers. Optimizing interface quality and tuning layer thicknesses and periodicity is a route toward simultaneously meeting the requirements for field/temperature stability of skyrmions as well as their velocity in devices, and the energy barriers to creation and deletion.

Toward these ends, multilayer thin films based on repeat units of [Pt/Co/Z], have attracted interest due to the out-of-plane anisotropy of Co and the additive DMI generated by choice of element $Z$ in the asymmetric Z/Co and Pt/Co interfaces [4–7]. Skyrmions in these structures have been observed using a variety of techniques including magnetic force microscopy (MFM) [6, 8, 9], x-ray magneto-circular dichroism (XMCD) [4, 6, 10], magneto-optical Kerr effect (MOKE) [11], and Lorentz transmission electron microscopy (LTEM) [9, 12, 13]. Various choices of $Z$ have been explored theoretically to better understand the impact on magnetic properties such as interfacial DMI, interlayer coupling (e.g. ferro- and anti-ferromagnetic), magnetic anisotropy, and thus the stability and size of skyrmions [14]. Focusing on the Pt/Co/Cu system, nanoscale skyrmions stable at zero applied magnetic field were predicted, with a size that is particularly sensitive to Co-layer thickness in the thin film regime [14]. Recently, our team has explored these predictions in a series of Pt/Co/Cu structures grown by molecular beam epitaxy [9, 15]. Bulk hysteresis loops measured with MOKE and SQUID exhibited a characteristic shape for samples within a certain range of layer thickness and repeat units. For these samples, LTEM and MFM imaging indicates a labyrinth phase at room temperature which transitions to a low density of ~ 130 nm Néel-type skyrmions with small, applied field (~ 0.1 T) and/or applied current pulses [9]. The pinning potentials which dictate skyrmion motion in such measurements are not well understood, in part because the spatial resolution of most magnetic imaging techniques in practice is typically > 10 nm.

This motivates our studies of Pt/Co/Cu thin films using spin-polarized scanning tunneling microscopy (SP-STM), which can simultaneously image magnetic textures and local structure with sub-nanometer resolution. While SP-STM has played a leading role exploring skyrmions in epitaxial magnetic monolayers on single crystal substrates [16–18], it has not previously been used to study thicker magnetic multilayer samples that are better suited for applications. Our STM imaging indicates the surface of these films comprises a connected network of nanoscale grains. The surface structure is highly uniform, with ~ 0.5 nm surface roughness over micron-scale areas. SP-STM imaging with a ferromagnetic Ni tip at 80 K reveals a variety of irregularly-shaped bubble textures, with effective diameters ranging from ~20-200 nm. Given the DMI-induced chirality revealed in LTEM [9, 15], analysis of the SPSTM contrast across the bubbles with different tip terminations leads us to conclude these textures are skyrmion bubbles. Close comparison of topographic and differential conductance (dI/dV) images reveals no apparent correlation between the skyrmion bubbles and the much smaller surface grain structure. The skyrmion bubbles can be nucleated or shifted by current/voltage pulses and motion along the surface is also influenced by the stray field of the tip. These studies suggest that pinning sites in such samples may be defined by subtle variations in film structure, or perhaps interface quality in sub surface layers within the heterostructure.

The [Pt/Co/Cu]$_n$ multilayer film studied here was grown by molecular beam epitaxy (MBE) in a chamber with base pressure better than 2×10$^{-9}$ Torr. MOKE and SQUID magnetometry of these samples indicated out-of-plane anisotropy, with $T_C$ > 300 K, and rounded hysteresis loops reflecting noncollinear magnetic textures [9]. For STM studies, a 5 nm Pt contact pad was first deposited onto a corner of the Al$_2$O$_3$(0001) substrate through a shadow mask. A Ta clip was then positioned onto the pad *ex situ*, before the substrate was brought back into UHV for the film growth. In our experience, this procedure makes reliable electrical contact to thin films grown on insulating substrates for STM studies. A 5nm Pt(111) buffer layer was then epitaxially grown on the Al$_2$O$_3$(0001) substrate using the following recipe: (1) deposit 3 atomic layers of Pt at sample temperature T = 450 $^\circ$C, (2) deposit the rest of the Pt layer while passively cooling down to T = 140 $^\circ$C, and (3) anneal at T = 300 $^\circ$C for 10 minutes. The [Pt/Co/Cu]$_n$ multilayers were then deposited on top of the buffer layer at room temperature. The thickness of each layer was controlled by the growth rates and the timing of the cell shutters. The deposition rates were measured prior to the growth by a quartz crystal monitor calibrated by x-ray reflection. The sample studied here is a Pt (5 nm)/[Co (1.2 nm)/Cu (0.4 nm)/Pt (0.45 nm)]$_4$/Co(1.2 nm) multilayer film that is shown schematically in Figure 1a. The film was terminated with the 1.2 nm Co layer to facilitate SP-STM, which is inherently most sensitive to the surface magnetism [19].

Following growth, the sample was transferred with a Ferrovac UHV suitcase to the load lock of a homebuilt UHV system with a 'bolt-on' RHK microscope head. No further surface preparation methods (e.g., Ar$^+$ ion sputtering, annealing) were used, to preserve the as-grown quality of the film surface. Guided by LTEM/MFM imaging showing skyrmion nucleation with magnetic field [9, 15], the sample was magnetized out-of-plane by holding it close to a permanent NdFeB magnet at room temperature before being transferred to the cold STM at ~80 K. All SP-STM measurements were performed at 80 K using a ferromagnetic, bulk Ni tip which was electrochemically etched [20–22] down to an apex diameter less than ~ 1 micron. The Ni tip was also magnetized anti-parallel to the sample's magnetization before being transferred into the cold STM. SP-STM dI/dV signals were obtained by adding a modulation voltage (50 mV$_{rms}$, 750 Hz) to the DC sample bias, and using a lock-in amplifier to measure the resultant dI/dV signal.

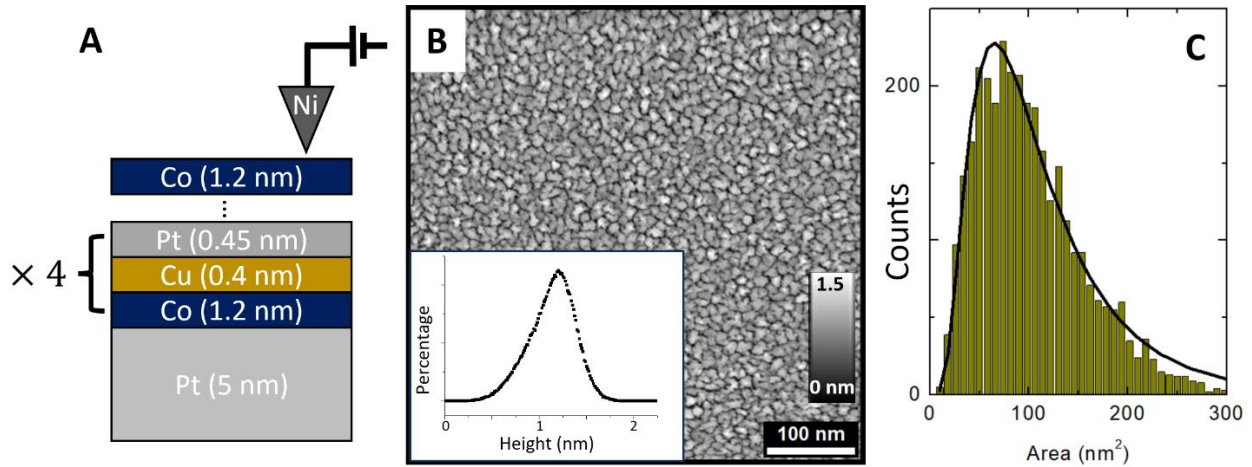

**Figure 1. Characterization of surface grain structure.** (A) Schematic of the multilayer sample. (B) STM topographic image of the surface structure. The inset shows ~ 0.5 nm surface roughness. (0.5 V, 200 pA) (C) Histogram of grain areas, showing a distribution peaked around ~ 70 nm$^2$. The solid curve is a fit to a log-normal distribution.

Topographic and dI/dV images were acquired simultaneously at fixed voltage while the feedback loop was engaged. Image processing and analysis was performed using *Gwyddion* [23] and SPIP. All topographic images were flattened to correct for scan-related bowing. Differential conductance images do not require flattening, and are presented with only some Gaussian smoothing for clarity.

A topographic STM image of the multilayer film surface is shown in Figure 1b. The surface structure consists of a dense network of closely packed grains. The film is highly uniform over the micron-scale areas imaged throughout the experiments, and we never observed distinct atomic steps related to the sapphire substrate, nor indication of spatial variation in the layer thicknesses. Thus, the primary source of the surface roughness is related to the individual grains, and a histogram of the surface topography reveals a variance in apparent height of ~ 0.5 nm (inset). This is a fraction of the nominal 1.2 nm Co layer termination, which suggests that the surface is uniformly terminated with Co. To analyze the statistics of the lateral area of these grains, we used the particle analysis functionality within *Gwyddion* [23] to identify and measure the thousands of grains within the images. The histogram in Fig. 1c peaks at ~ 70 nm$^2$, with only a small tail of the distribution extending to larger areas (> 250 nm$^2$).

Figure 2 compares a micron-scale STM topographic image with a simultaneously acquired dI/dV image. While the topographic image shows that the uniform granular structure persists over this larger scale, the dI/dV image shows a dark background decorated with bright bubbles. The uniformly dark background is consistent with anti-alignment of the sample and tip magnetizations out-of-plane. The increased dI/dV signal of the bubbles indicates the expected alignment of the core parallel to the tip's magnetization and anti-parallel to the ferromagnetic background, as the SP-STM tunneling magnetoresistance (TMR) contribution varies as $\cos(\theta)$, where $\theta$ is the angle between the tip and sample magnetizations [19]. Although the bubbles can be somewhat irregularly shaped, we performed particle analysis to estimate the size distribution, and we find that the skyrmions have effective diameters ranging from 20-200 nm. This range is consistent with the skyrmions observed with MFM and Lorentz TEM at room temperature [9, 15], and measurements of skyrmions in similar magnetic multilayers [4, 6, 24].

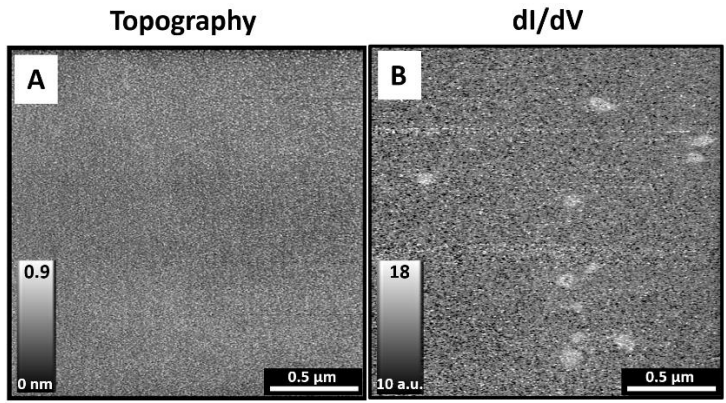

**Figure 2. SP-STM images of skyrmions.** (A) STM topographic image (B) Simultaneously acquired SP-STM dI/dV image showing bright contrast associated with the skyrmions. (0.3 V, 50 pA)

To explore the correlation of structure and magnetism in more detail, Figure 3 shows higher magnification topographic and dI/dV images of one bubble. While the granular structure is again evident in the topographic image, it is largely absent in the dI/dV image, consistent with a uniformly magnetized, cobalt-terminated surface. Shown below the images are corresponding line profiles of the topographic and dI/dV signals. While the topography is flat within the height variations of the granular structure, the dI/dV profile shows an abrupt increase in signal at the bubble edges (within ~ 10 nm), and a relatively flat plateau of magnetic contrast across the ~ 150 nm bubble. Previously, LTEM data on these samples revealed a uniform bubble chirality, set by the interfacial DMI [9]. These two observations suggest these spin textures are skyrmion bubbles, which can be thought of as wrapping of a Néel-type domain wall around a uniformly magnetized core, rather than a pure skyrmion where a continuous winding of magnetization from edge to center is expected. Skyrmion bubbles share the topological character of skyrmions, and the uniformly magnetized core of larger trivial bubbles [25]. While the bubbles in similar samples were imaged with nearly circular contrast in previous LTEM and MFM studies [9, 15], the higher spatial resolution of SPSTM clearly shows bubbles with irregular shapes. The dashed outline in the topographic image corresponds to the skyrmion bubble boundary in the dI/dV image, and we find no apparent correlation between local variations in the surface structure and the skyrmion bubble size or shape. Thus, the variation in skyrmion bubble morphology in Figure 2b likely reflects a sensitivity to subtle changes in the underlying Co, Pt or Cu layers, or aspects of the metal/$Al_2O_3$ interface that may not be directly probed by SP-STM imaging of the surface. These structural variations could influence underlying micromagnetic parameters such as pinning potential, exchange stiffness, interfacial DMI, or local anisotropy.

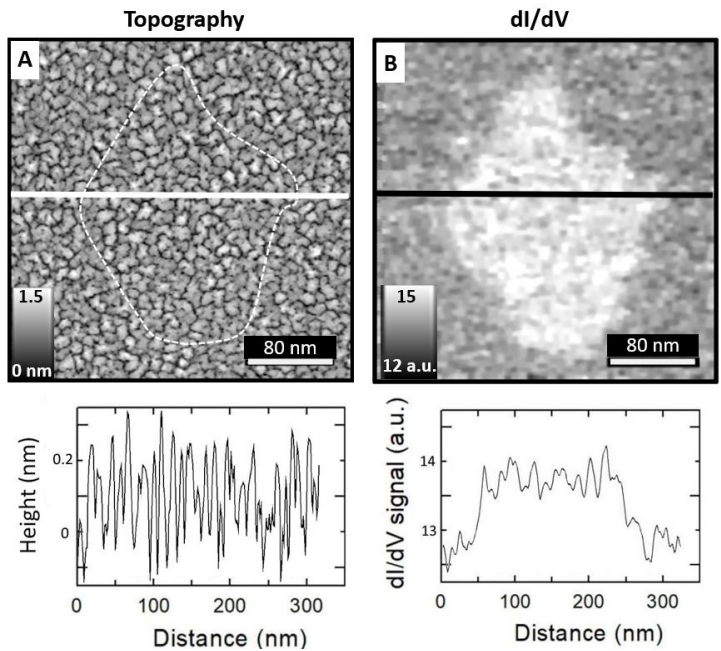

Figure 3. Comparison of topography and SP-STM at high magnification. (A) STM topographic image. The dotted white line indicates the outline of the skyrmion bubble revealed in the simultaneously acquired dI/dV image (B). Below each image are corresponding line profiles. (0.3 V, 50 pA).

While SP-STM imaging of the skyrmion bubbles using the Ni tip was reproducible and reliable over many days of imaging, we did observe occasional, random changes in the tip termination which provide additional insights. Such changes in STM experiments generally reflect a rearrangement of atoms on the tip apex due to the dropping off or picking up of clusters, molecules or impurity atoms (e.g. carbon, oxygen). In the STM topographic image of Figure 4a, a random change in tip apex produces a sudden 2 nm step down at the scan line indicated by the arrow and the granular structure is less sharply defined in subsequent imaging. The surface was not apparently perturbed however, suggesting that only the atoms at the tip apex were rearranged. The dI/dV image in Figure 4b shows an abrupt change in the image contrast of the skyrmion bubbles to a core whose contrast is indistinguishable from the background, along with a bright halo around the perimeter. Similar donut-like contrast was reported using nonmagnetic tips in previous SP-STM studies of nanoscale skyrmions in Rh/Co bilayer films and discussed with respect to tunneling anisotropic- or noncollinear magnetoresistance mechanisms (TAMR, NCMR) [16, 26, 27]. While TAMR reflects the tunneling probably into regions with perpendicular spin alignments, the NCMR mechanism reflects an evolution of the local density of states throughout a canted spin texture. For the skyrmion bubbles here, noncollinearity of the spins is concentrated in the domain walls separating the antiparallel core spins from the ferromagnetic background. Consistent with Figure 4, the TAMR and NCMR mechanisms would produce contrast localized to the domain wall region, with zero contrast in both the ferromagnetic background and core region. We estimate a wall width ~ 30-40 nm, which is consistent within experimental error with our prior LTEM imaging in the labyrinth phase [9]. Further study comparing magnetic-field and spatially-resolved spectroscopy would be needed to distinguish between these possible mechanisms. We note that a canted tip with in-plane magnetization would lead to a more double-lobed appearance of these Néel-type skyrmion bubbles [28], in contrast to the observation here, suggesting that the tip termination in Figure 4 is non-magnetic.

Nucleation, annihilation, and motion of skyrmions due to external stimuli is important for future applications, and is subject to a landscape of magnetic pinning potentials that is still not well understood. In our SP-STM studies of these Co/Pt/Cu multilayer samples, we have observed numerous examples of skyrmion bubbles being nucleated with the current/voltage pulses used to sharpen the SP-STM tip,

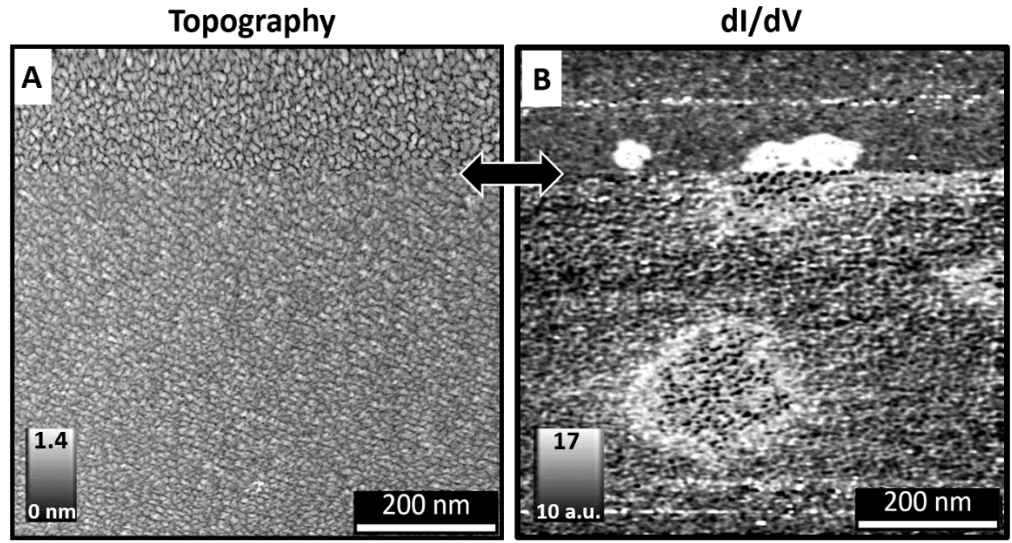

**Figure 4. Change in dI/dV imaging contrast due to NCMR or TAMR.** (A) STM topographic image where a change in the tip termination occurred during the scan line indicated by the black arrow. A 2 nm step down in apparent height was subtracted from the image for clarity. (B) Corresponding dI/dV image, showing a change from uniformly bright contrast to a more halo-like contrast consistent with an NCMR or TAMR mechanism as described in the text. (0.3 V, 50 pA)

skyrmion bubbles changing shape, and skyrmion bubble annihilation. Figure 5 illustrates these effects in a series of dI/dV images of the same area, taken over a course of 8 hours. Initially, a single skyrmion bubble was observed in this area (Fig. 5a), but two more were nucleated during scanning, evidenced by the abrupt onset of bright contrast within one scan line in Fig. 5b. The original skyrmion bubble has also shifted to the right, and is somewhat larger. In the next frame (Fig. 5c), the original skyrmion bubble shifts further to the right, while the left skyrmion bubble appears pinned in place. In Figures 5d-e, the smaller of the nucleated skyrmion bubbles appears to have merged with the original one, which also appears elongated as it continues to move to the right. The left skyrmion bubble remains fixed in place but does get a little smaller. Lastly, in Figure 5f, the original skyrmion bubble has shifted out of the field of view to the right of the image, while the left skyrmion bubble has annihilated and is no longer observed. Though it is difficult to rule out thermal effects, at least some of this motion appears to be induced by the STM tip during scanning, evidenced by abrupt changes within one scan line as the tip is brought near the skyrmion bubbles. Because we are using a bulk Ni tip, it is likely that the stray field of the tip is influencing the skyrmion bubble's motion, which is consistent with the relatively small nucleation field (100 mT) for bubbles in these samples [9, 15].

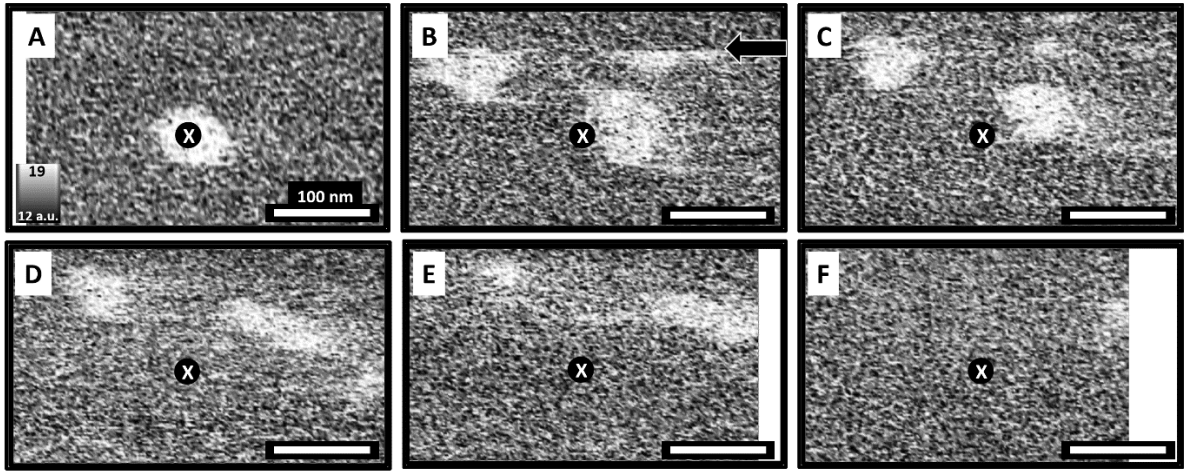

**Figure 5. Nucleation, annihilation, and motion of skyrmion bubbles.** (A-F) Sequence of dI/dV images of the same topographic area. The 'X' marks a fixed reference point taken from the corresponding topographic images (not shown). The arrow in (B) indicates the sudden appearance of two skyrmion bubbles near the top of the image. Motion and annihilation in subsequent images likely is influenced by the stray field of the Ni SP-STM tip. The slow scan direction was downward and the fast scan direction was left/right in the images. (0.3 V, 60 pA)

In conclusion, these studies represent an initial effort to correlate surface structure and magnetism using SP-STM in magnetic multilayers. We observe a range of skyrmion bubble sizes and pinning behaviors, even when separated by less than 100 nm. Our detailed comparison of STM topography and dI/dV images shows that there are no surface defects or inhomogeneities at length scales that could account for the range in skyrmion bubble sizes or variations in pinning. Further studies with varying temperature and magnetic field would help better quantify the relative energy scales underlying these variations. Comparing these SP-STM studies with more volumetric techniques such as LTEM can help disentangle the role of surface and sub-surface inhomogeneities, and thus help guide the optimization of these multilayers for potential applications.

**Acknowledgements -** Primary support was provided by DARPA Grant No. D18AP00008. Partial support was provided by DOE Grant DE-SC0016379 for data analysis and SP-STM technique development. **Author contributions:** J.R. and B.G. performed SP-STM experiments, analyzed data and wrote the manuscript, S.C. grew and characterized the samples, and R.K.K., and J.A.G. helped in writing the manuscript and interpreting the results. **Competing interests:** Authors declare no competing interests.